\documentclass[twocolumn,english]{revtex4-2}
\usepackage[T1]{fontenc}
\usepackage[utf8]{inputenc}
\usepackage{amsmath}
\usepackage{amssymb}
\usepackage{graphicx}
\usepackage{babel}
\begin{document}
\title{Effective Dynamics of Tracer in Active Bath: A Mean-field Theory Study }
\author{Mengkai Feng, Zhonghuai Hou}
\email{Corresponding Author: hzhlj@ustc.edu.cn}

\affiliation{Department of Chemical Physics \& Hefei National Laboratory for Physical
Sciences at the Microscale, iChEM, University of Science and Technology
of China, Hefei, Anhui 230026, China}
\begin{abstract}
We develop a theoretical framework to study the effective dynamics
of a tracer immersed in a nonequilibrium bath consisting of active
particles. By using a mean-field approximation and extending the linearized
Dean equation to nonequilibrium environment, we derive a generalized
Langevin equation for the tracer particle, wherein colored noise terms
and a memory kernel reflect the roles of interactions between tracer
and bath particles as well as activity of the bath. In particular,
we obtain a self-consistent equation to calculate the long time diffusion
coefficient and mobility of tracer, finding that they both increase
non-linearly with bath activity, in good consistents with direct simulation
results. 
\end{abstract}
\maketitle

\section{Introduction}

Tracer dynamics in nonequilibrium baths have attracted great attention
in recent decades\citep{2000_PRL_Wu_bacteria_bath,2010_EPL_Seifert_Speck_FDT,15_PRE_Maes_FricNoiseProbe,2016_RMP_BechingerActive,2018_NatComm_Bechinger_probe,19_JCP_Brady_FDT,2020_PRL_YangMC},
which is an important topic for understanding many biological processes
and artificial active particles systems\citep{2000_PRL_Wu_bacteria_bath,2007_PRL_Chen_FlucRheoActiSuspen,2011_RPL_Angelani_EffecInteCollParti,2014_PRL_Maggi_GenaralEnergyEquiInAB,2017_SciRep_Maggi_MemLessResponseAndFDT}.
For instance, complex intracellular environment can be abstracted
as an active bath, since the molecular motors in cytoplasm are similar
to active particles. Consequently, biomolecules in living cells often
show anomalous behaviors (such as anomalous diffusion) due to this
nonequilibrium feature\citep{2002_PRL_AnomoDiff}. Besides, the presence
of active fluctuations may promote protein folding, which has been
observed in a polymer model system\citep{14_PRE_ActPolym}. On the
other hand, the passive tracers in a bath of swimming bacteria show
novel properties such as superdiffusion\citep{2000_PRL_Wu_bacteria_bath},
effective attractive interactions\citep{2011_RPL_Angelani_EffecInteCollParti},
and even targeted delivery\citep{13_NC_TargetDelivery}. Because of
the importance and wide range of applications, developing a general
theoretical framework to describe the dynamics of tracer in such nonequilibrium
baths is desirable.

In this regard, some important progresses have been made in recent
years\citep{2007_JCP_DDFT,2009_PRL_Maes,2011_PRE_Demery_TracerInFluids,2011_JSM_Maes_FluRes,2013_PRE_Maes,2014_NJP_Demery_GLEforDrivenTracer,2014_PRE_Szamel_EffecTemp,2016_JPCM_Maes_Langevin,2017_PRE_TracerDiffActBath,20_PRL_Maes,20_FrontPhys_Maes_respon},
including density functional theory\citep{2007_JCP_DDFT}, nonequilibrium
linear response theory\citep{2017_PRE_TracerDiffActBath,2009_PRL_Maes,2013_PRE_Maes,2011_JSM_Maes_FluRes,2016_JPCM_Maes_Langevin,20_PRL_Maes,20_FrontPhys_Maes_respon},
mean-field theory\citep{2014_NJP_Demery_GLEforDrivenTracer,2011_PRE_Demery_TracerInFluids,2019_JSM_Dean_DrivenProbeInHarmoConfi,2019_PRL_Demery_SSPinHarmoTrap,2010_PRL_Demery_DragForce},
and even mode-coupling theory\citep{09_PRL_Fuchs,2013_PRE_Fuchs}.
For instance, Speck and Seifert \textit{et al} established a fluctuation-dissipation
theorem (FDT) in nonequilibrium steady state of sheared colloidal
suspension system\citep{2010_EPL_Seifert_Speck_FDT,2011_EPL_SeifertSpeck_MobDiff},
then studied the mobility and diffusivity of a tagged particle in
such system\citep{2011_EPL_SeifertSpeck_MobDiff}. Starting from a
different path, Brady \textit{et al} used Smoluchowski equation to
investigate the long-time diffusivity of a tracer submerged in active
suspensions\citep{2017_PRE_TracerDiffActBath,19_JCP_Brady_FDT}, and
further derived a general relationship between diffusivity and mobility
by testing the results with Stokes-Einstein-Sutherland relation\citep{19_JCP_Brady_FDT}.
In the fundamental respect, Maes \textit{et al} obtained a generalized
fluctuation-dissipation relation (FDR) for the linear response of
driven systems\citep{2009_PRL_Maes,2011_JSM_Maes_FluRes}, studied
the effective dynamics of Brownian particle with strong interaction
with the medium, and extended this FDR to non-equilibrium systems
with memory \citep{2013_PRE_Maes}. Most recently, they used their
method to derive the fluctuation dynamics of a probe in weak coupling
with an active environment such as living tissue\citep{20_PRL_Maes}. 

In the present article, we propose an alternative theoretical method
based on mean-field theory to study the effective dynamics of tracer
in active bath. Starting from the Langevin equations (LEs) for the
whole system, we achieve the linearized Dean's equation \citep{2011_JPCM_Dean_DiffActTracInFlucField}
for the evolution of density fluctuation of bath particles, wherein
the interactions between tracer and bath particles are involved explicitly.
By substituting the density fluctuation into the tracer's LE, we derive
a generalized Langevin equation (GLE) for tracer, which includes a
memory kernel function and complex effective noise terms, of which
the correlations of such noises are strongly coupled with tracer dynamics.
This GLE facilitates us to calculate the effective diffusion and mobility
of the tracer. With an appropriate approximation, these transport
coefficients can be conveniently calculated through self-consistent
equations, showing good agreements with direct molecular dynamics
simulation data.  

This article is organized as following. In Sec.II we derive the Dean's
equation for bath density fluctuation and the GLE for tracer. In Sec.III
we use the GLE to calculate the effective diffusion coefficient and
mobility of the tracer, and compare them with simulation results.
Details of the derivations are shown in Appendix \ref{sec:detail}.

\section{Model and Theory}

\subsection{Equations of Motion}

\begin{figure}
\begin{centering}
\includegraphics[width=6cm]{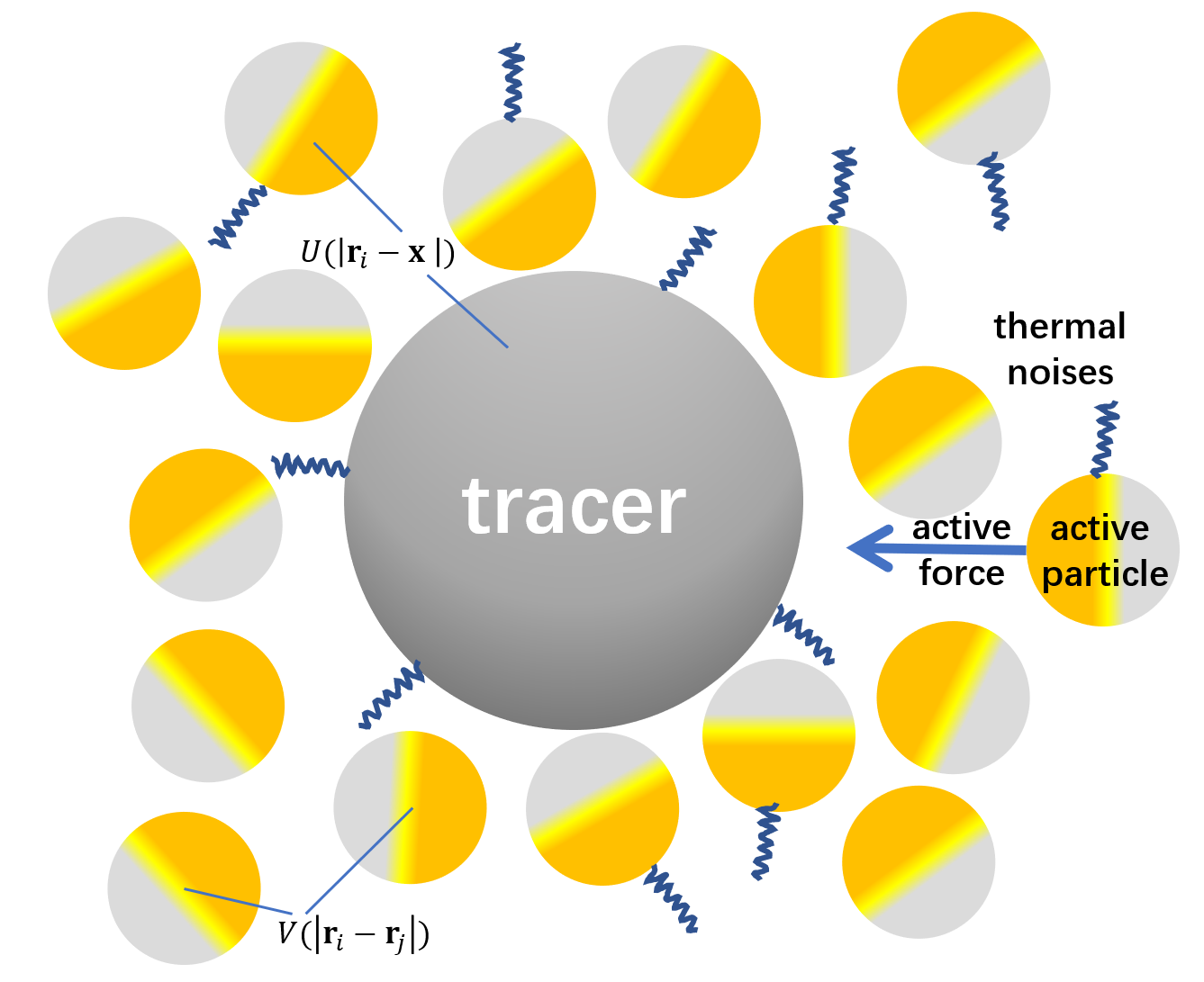}
\par\end{centering}
\caption{Schematic figure of the model system. The bath particles are active
and the tracer is passive and the size is larger than that of the
bath particles. Both of them are immersed in the background equilibrium
reservoir, therefore influenced by thermal white noises.}
\label{fig:act_bath}
\end{figure}

In the present work, we consider a two-dimensional (2D) system composed
of a spherical passive tracer of radius $R_{tr}$ surrounded by $N$
self-propelled active bath particles of radius $R_{b}$, both in a
background equilibrium thermal bath with temperature $T$. The tracer
position is denoted by ${\bf x}$ and those of the bath particles
are given by ${\bf r}_{i=\left(1,...N\right)}$. The dynamics of the
tracer are described by the overdamped Langevin equation 
\begin{equation}
{\rm d}{\bf x}=-\mu_{t}\nabla_{{\bf x}}\sum_{i}U\left(\left|{\bf r}_{i}-{\bf x}\right|\right){\rm d}t+\sqrt{2\mu_{t}T}{\rm d}{\bf W}_{x},\label{eq:TracerLE}
\end{equation}
where $\mu_{t}$ is the bare mobility of tracer, $U\left(\left|{\bf r}_{i}-{\bf x}\right|\right)$
is the pair interaction potential between the tracer and bath particle
$i$, ${\bf W}_{x}$ denotes a standard Wiener process accounting
for thermal noise such that $dW_{x}\left(t\right)=\xi\left(t\right)dt$
with $\left\langle \xi\left(t\right)\right\rangle =0$ and $\left\langle \xi\left(t\right)\xi\left(t'\right)\right\rangle =\delta\left(t-t'\right)$.
For the bath particles, we model the self-propulsion force by the
Ornstein-Uhlenbeck (OU) noise \citep{19_PRE_Bonilla_AOUP}, therefore
the dynamics are given by 
\begin{align}
{\rm d}{\bf r}_{i}= & -\mu_{b}\nabla_{i}\left[\sum_{j\neq i}V\left(\left|{\bf r}_{i}-{\bf r}_{j}\right|\right)+U\left(\left|{\bf r}_{i}-{\bf x}\right|\right)\right]{\rm d}t\nonumber \\
 & +{\bf f}_{i}{\rm d}t+\sqrt{2\mu_{b}T}{\rm d}{\bf W}_{i},\label{eq:LE_Bath}
\end{align}
where $\mu_{b}$ is the bare mobility of bath particles, $V\left(\left|{\bf r}_{i}-{\bf r}_{j}\right|\right)$
is the interacting potential between bath particles $i$ and $j$,
and ${\bf W}_{i}$ accounts for thermal noise for particle $i$. Active
force ${\bf f}_{i}$ is the OU-colored noise type, governed by 
\begin{equation}
\tau_{b}{\rm d}{\bf f}_{i}=-{\bf f}_{i}{\rm d}t+\sqrt{2D_{b}}{\rm d}{\bf W}_{i}^{f},\label{eq:OUnoise}
\end{equation}
where $\tau_{b}$ denotes the persistent time of self-propulsion,
$D_{b}$ is an equivalent diffusion coefficient, and ${\bf W}_{i}^{f}$
is also a standard Wiener process. The time correlation of ${\bf f}_{i}$
is then given by 
\begin{equation}
\left\langle {\bf f}_{i}\left(t\right){\bf f}_{j}\left(t'\right)\right\rangle =\frac{D_{b}}{\tau_{b}}\delta_{ij}e^{-\left|t-t'\right|/\tau_{b}}{\bf I}
\end{equation}
wherein $i(j)$ stands for particle label and ${\bf I}$ is unit
tensor. The model used here for the active particles is known as the
AOUP model. Note that in many studies\citep{2015_Annu_Cates_MIPS,2016_RMP_BechingerActive,20_ActiveMatter},
the active Brownian particle(ABP) model is also often used, wherein
the active force is realized via a constant velocity $v_{0}$ along
a random rotational direction given by ${\bf e}_{\varphi}=\left(\cos\varphi,\sin\varphi\right)$,
where $\left\langle \dot{\varphi}\left(t\right)\dot{\varphi}\left(t'\right)\right\rangle =2D_{r}\delta\left(t-t'\right)$
with $D_{r}$ the rotational diffusion coefficient. It can be shown
that the correlation function of the active force of ABP model at
a coarse-grained time scale is the same in form as that of AOUP model.
To better illustrate our model, we have shown a schematic diagram
of the system in Fig. \ref{fig:act_bath}.

\subsection{Density Fluctuation of Active Bath: Mean field approximation }

The goal of the present work is to obtain an effective (reduced) equation
of motion for the tracer particle, by tracing out the degrees of freedom
of the bath particles. To this end, one needs to compute the force
exerted by the bath particles on the tracer in a mean-field manner.
Following the procedures as described in \citep{1996_JPA_Dean_LE},
we first describe the bath dynamics at a coarse-grained level, by
introducing the bath density field $\rho\left({\bf r},t\right)=\sum_{i}\rho_{i}\left({\bf r},t\right)$
where $\rho_{i}({\bf r},t)=\delta\left({\bf r}-{\bf r}_{i}\left(t\right)\right)$,
and using standard techniques \citep{1996_JPA_Dean_LE} to derive
the time evolution equation (see appendix for details), \onecolumngrid
\begin{align}
\frac{\partial\rho\left({\bf r},t\right)}{\partial t}= & \mu_{b}\nabla\cdot\left\{ \rho\left({\bf r},t\right)\Big[\int{\rm d}{\bf r}'\rho\left({\bf r}',t\right)\nabla V\left(\left|{\bf r}-{\bf r}'\right|\right)+\nabla U\left(\left|{\bf r}-{\bf x}\right|\right)\Big]\right\} \nonumber \\
 & +\mu_{b}T\nabla^{2}\rho\left({\bf r},t\right)+\nabla\cdot\left(\sqrt{\rho\left({\bf r},t\right)}\boldsymbol{\xi}^{T}\right)+\nabla\cdot\left(\sqrt{\rho\left({\bf r},t\right)}\boldsymbol{\xi}^{A}\right)\label{eq:Dean}
\end{align}
\twocolumngrid \noindent where $\boldsymbol{\xi}^{T}$ and $\boldsymbol{\xi}^{A}$
are spatiotemporal Gaussian noises, with correlations 
\begin{equation}
\left\langle \boldsymbol{\xi}^{T}\left({\bf r},t\right)\boldsymbol{\xi}^{T}\left({\bf r}',t'\right)\right\rangle =2\mu_{b}T\delta\left(t-t'\right)\delta\left({\bf r}-{\bf r}'\right){\bf I}\label{eq:Corr-Xi-T}
\end{equation}
 and 
\begin{equation}
\left\langle \boldsymbol{\xi}^{A}\left({\bf r},t\right)\boldsymbol{\xi}^{A}\left({\bf r}',t'\right)\right\rangle =\frac{D_{b}}{\tau_{b}}e^{-\left|t-t'\right|/\tau_{b}}\delta\left({\bf r}-{\bf r}'\right){\bf I}\label{eq:Corr-Xi-A}
\end{equation}
 respectively. 

On the right-hand-side of Eq.(\ref{eq:Dean}), the first term describes
how the interactions $U$(particle-tracer) and $V$(particle-particle)
would influence the density field. The second term is a diffusive
one that accounts for the background thermal noise at temperature
$T$. In the third term, $\boldsymbol{\xi}^{T}$ ('T' for thermal)
denotes a coarse-grained white noise, which also originates from the
thermal background, coupled with the density field itself. $\boldsymbol{\xi}^{A}$
('A' for active) in the fourth term denotes a coarse-grained OU-noise
resulting from the coupling between the active forces ${\bf f}_{i}$
and the density field, which would be absent for a passive bath. 

To proceed further, we assume that the system is nearly homogeneous(i.e.
no phase separation or dynamic heterogeneous) and thus the bath dynamics
is mainly dependent on the density fluctuation $\delta\rho\left({\bf r},t\right)=\rho\left({\bf r},t\right)-\bar{\rho}$,
where $\bar{\rho}$ is the number density of bath particles. Using
the methods in Ref.\citep{2011_JPCM_Dean_DiffActTracInFlucField,2014_NJP_Demery_GLEforDrivenTracer}(also
shown in Appendix \ref{subsec:Deri_dean}), on the condition of weak
interaction and high density, the mean-field approximation is applicable
and one can solve the linearized equation of bath density fluctuation
in Fourier space
\begin{align}
\partial_{t}\delta\rho_{k}(t)\approx & -\mu_{b}k^{2}\left[\bar{\rho}U_{k}e^{i{\bf k}\cdot{\bf x}}+\left(T+\bar{\rho}V_{k}\right)\delta\rho_{k}(t)\right]\nonumber \\
 & +\sqrt{\bar{\rho}}i{\bf k}\cdot\left[\tilde{\boldsymbol{\xi}}_{k}^{T}\left(t\right)+\tilde{\boldsymbol{\xi}}_{k}^{A}\left(t\right)\right]\label{eq:dean_flu}
\end{align}
where $\delta\rho_{k}\left(t\right)=\int\delta\rho\left({\bf r},t\right)e^{i{\bf k}\cdot{\bf r}}{\rm d}{\bf r}$
is the Fourier mode of density fluctuation $\delta\rho\left({\bf r},t\right)$
(isotropic condition has been used such that $\delta\rho_{{\bf k}}$
only depends on $k=\left|{\bf k}\right|$), and $V_{k}$ , $U_{k}$,
$\tilde{\boldsymbol{\xi}}_{k}^{T}\left(t\right)$ and $\tilde{\boldsymbol{\xi}}_{k}^{A}\left(t\right)$
are Fourier modes defined in the same manner for $V\left(r\right)$,
$U\left(r\right)$, $\boldsymbol{\xi}^{T}\left(t\right)$ and $\boldsymbol{\xi}^{A}\left(t\right)$
respectively. The formal solution of \eqref{eq:dean_flu} is then
given by 

\begin{multline}
\delta\rho_{k}(t)\approx\int_{-\infty}^{t}{\rm d}se^{-(t-s)/a_{k}}\left[-\mu_{b}k^{2}\bar{\rho}U_{k}e^{i{\bf k}\cdot{\bf x}\left(s\right)}\right.\\
\left.+i\sqrt{\bar{\rho}}{\bf k}\cdot\left(\tilde{\boldsymbol{\xi}}_{k}^{T}\left(s\right)+\tilde{\boldsymbol{\xi}}_{k}^{A}\left(s\right)\right)\right],\label{eq:dean_solu}
\end{multline}
where $a_{k}=\left[\mu_{b}k^{2}\left(T+\bar{\rho}V_{k}\right)\right]^{-1}$
is a characteristic relaxation time scale depending on $k$ and the
interaction potential $V$ among bath particles. Generally, $a_{k}$
scales as $k^{-2}$ so that fluctuations with short wavelength decays
very fast. 

\subsection{Generalized Langevin Equation of Tracer}

Now we turn to the tracer dynamics. Using the identity(proceeding
the Fourier transform twice)

\begin{equation}
\nabla_{x}\sum_{i}U\left(\left|{\bf r}_{i}-{\bf x}\right|\right)=-\int_{{\bf k}}i{\bf k}e^{-i{\bf k}\cdot{\bf x}}\delta\rho_{k}U_{k}
\end{equation}
where $\int_{{\bf k}}=\left(2\pi\right)^{-d}\int{\rm d}k^{d}$ is
the integral over entire k-space, and inserting the formal solution
of $\delta\rho_{k}$, Eq.\eqref{eq:dean_solu}, into Eq.\eqref{eq:TracerLE},
we can obtain a generalized Langevin equation (GLE) of the tracer
\begin{align}
\dot{{\bf x}}\left(t\right)= & \mu_{t}\int_{-\infty}^{t}{\bf F}\left({\bf x}\left(t\right)-{\bf x}\left(s\right),t-s\right){\rm d}s\nonumber \\
 & +\boldsymbol{\eta}_{A}\left({\bf x}\left(t\right),t\right)+\boldsymbol{\eta}_{T}\left({\bf x}\left(t\right),t\right)+\sqrt{2\mu_{t}T}\boldsymbol{\xi}_{t},\label{eq:GLE1}
\end{align}
where 
\begin{align}
 & {\bf F}\left({\bf x}\left(t\right)-{\bf x}\left(s\right),t-s\right)\nonumber \\
= & -\mu_{b}\bar{\rho}\int_{{\bf k}}i{\bf k}k^{2}U_{k}^{2}e^{-i{\bf k}\cdot\left[{\bf x}\left(t\right)-{\bf x}\left(s\right)\right]}e^{-\left(t-s\right)/a_{k}}
\end{align}
is a memory kernel term, $\boldsymbol{\xi}_{t}={\rm d}{\bf W}_{x}\left(t\right)/{\rm d}t$
is the original thermal noise of the background equilibrium bath upon
the tracer, and $\boldsymbol{\eta}_{A,T}\left({\bf x}(t),t\right)$
are complicated colored noises that account for the effects of active
bath particles. In other words, the total force exerted by the bath
particles on the tracer now is decomposed into a systematic frictional
force with memory and a noise term, similar to that of a Brownian
particle in a thermal bath. Nevertheless, now the noise terms $\boldsymbol{\eta}_{A,T}\left({\bf x}(t),t\right)$
are much more complicated than a simple Gaussian white one, 

\begin{align}
\boldsymbol{\eta}_{A,T}\left({\bf x}\left(t\right),t\right)= & -\mu_{t}\int_{{\bf k}}{\bf k}e^{-i{\bf k}\cdot{\bf x}_{t}}U_{k}\int_{-\infty}^{t}e^{-\left(t-s\right)/a_{k}}\nonumber \\
 & \times\left[\sqrt{\bar{\rho}}{\bf k}\cdot\tilde{\boldsymbol{\xi}}_{k}^{A,T}\left(s\right)\right]{\rm d}s\label{eq:Eta_AT}
\end{align}
originated further from the coarse-grained thermal noise $\boldsymbol{\xi}^{T}$
or the coarse-grained active OU-noise $\boldsymbol{\xi}^{A}$, respectively
(the correlation functions of $\boldsymbol{\xi}^{T}$ and $\boldsymbol{\xi}^{A}$
are given in Eq.\eqref{eq:Corr-Xi-T} and Eq.\eqref{eq:Corr-Xi-A},
and $\boldsymbol{\eta}_{A,T}(t)$ can be seen as weighted sum of $\tilde{\boldsymbol{\xi}}_{k}^{A,T}\left(s\right)$
for all $s<t$). It is $\boldsymbol{\eta}_{A}\left({\bf x}\left(t\right),t\right)$
that accounts for the active nature of the bath, which becomes zero
when activity is absent. Note that the structure of Eq.\eqref{eq:Eta_AT}
is complicated, such that the properties of $\boldsymbol{\eta}_{A,T}$
are quite non-trivial. 

Keeping to the lowest order, we can approximately write the memory
kernel ${\bf F}\left({\bf x}\left(t\right)-{\bf x}\left(s\right),t-s\right)$
into a more familiar form as 
\begin{align}
\int_{-\infty}^{t}\mu_{t}{\bf F}\left({\bf x}\left(t\right)-{\bf x}\left(s\right),t-s\right){\rm d}s & \simeq-\int_{-\infty}^{t}{\rm d}u\zeta(t-u)\dot{{\bf x}}(u)
\end{align}
(see Appendix \ref{subsec:gle} for details), wherein the usual memory
kernel $\zeta\left(t\right)$ reads,
\begin{equation}
\zeta(t)=\frac{\mu_{t}\mu_{b}\bar{\rho}}{d}\int_{{\bf k}}k^{4}U_{k}^{2}a_{k}e^{-t/a_{k}}
\end{equation}
Consequently, the GLE for the tracer dynamics reads
\begin{equation}
\dot{{\bf x}}=-\int_{-\infty}^{t}\zeta(t-u)\dot{{\bf x}}(u){\rm d}u+\boldsymbol{\eta}_{A}\left(t\right)+\boldsymbol{\eta}_{T}\left(t\right)+\sqrt{2\mu_{t}T}\boldsymbol{\xi}_{t}\label{eq:GLE2}
\end{equation}
This equation is one of the main results of the present paper. It
allows us to calculate the transport properties of tracer in active
bath, such as mobility and diffusivity, and moreover to investigate
the nature of active bath itself. 

If one wants to investigate the long time behavior of the tracer,
such as diffusion dynamics, the memory effect in the GLE could be
ignored, such that one may adopt Markovian approximation as\citep{Dchandler_IntroMSM}
\begin{equation}
\int_{-\infty}^{t}\zeta(t-u)\dot{{\bf x}}(u){\rm d}u\simeq\left(\int_{0}^{\infty}\zeta(u){\rm d}u\right)\dot{{\bf x}}(t)\equiv\lambda\dot{{\bf x}}(t)
\end{equation}
with explicitly the effective friction coefficient $\lambda$ reads
\begin{align}
\lambda & =\int_{0}^{\infty}\zeta(u){\rm d}u=\frac{\mu_{t}\mu_{b}\bar{\rho}}{d}\int_{{\bf k}}k^{4}U_{k}^{2}a_{k}\nonumber \\
 & =\frac{\mu_{t}\bar{\rho}}{\mu_{b}d}\int_{{\bf k}}\frac{U_{k}^{2}}{\left(T+\bar{\rho}V_{k}\right)^{2}}.
\end{align}
Therefore, the GLE reduces to a simple Langevin equation(LE) with
colored noise 
\begin{equation}
\left(1+\lambda\right)\dot{{\bf x}}=\boldsymbol{\eta}_{A}\left(t\right)+\boldsymbol{\eta}_{T}\left(t\right)+\sqrt{2\mu_{t}T}\boldsymbol{\xi}_{t}\label{eq:LE}
\end{equation}
which serves as the working equation to study the diffusion dynamics
of the tracer. 

\subsection{Noise Correlations }

Previously, in most studies, the effect of the active bath on the
tracer is treated as an additive noise with Ornstein-Uhlenbeck (OU)
\citep{2000_PRL_Wu_bacteria_bath,2020_PRL_YangMC,2004_PF_Kim_EnhancedDiff,2014_PRL_Maggi_GenaralEnergyEquiInAB,2015_ScitiRep_Maggi_Multidimen_Stationary}
type, wherein the amplitude and relaxation parameter are obtained
through simulation or experimental data. Indeed, in certain aspects
such as long-time diffusion and spatial distribution of the tracer
in trapping potential, this approximation accords with the experimental
or simulation results. However, how interparticle interactions and
bath activity enter into such noise explicitly and how to derive the
expression for such complex effects are still unclear as far as we
know. In other words, the mechanism of noise generation is still unresolved
from a theoretical point of view. In this respect, our theory provides
an inspired understanding of this issue.

To investigate such noise properties, we use Eq.(\ref{eq:GLE2}) as
the starting point for further studies. Clearly, the dynamics of ${\bf x}$
strongly depends on the correlation property of the colored noises
$\boldsymbol{\eta}_{T}\left(t\right)$ and $\boldsymbol{\eta}_{A}\left(t\right)$.
After some straightforward calculations(see Appendix\eqref{subsec:actnoi}),
we have

\begin{align}
 & \left\langle \boldsymbol{\eta}_{T}\left(t\right)\cdot\boldsymbol{\eta}_{T}\left(t'\right)\right\rangle \nonumber \\
\approx & \mu_{t}^{2}\mu_{b}\bar{\rho}T\int_{{\bf k}}k^{4}U_{k}^{2}a_{k}\left\langle e^{-i{\bf k}\cdot\left[{\bf x}(t)-{\bf x}(t')\right]}\right\rangle e^{-\left|t-t'\right|/a_{k}}\label{eq:corr_eta_t}
\end{align}
and
\begin{align}
 & \left\langle \boldsymbol{\eta}_{A}\left(t\right)\cdot\boldsymbol{\eta}_{A}\left(t'\right)\right\rangle \approx\mu_{t}^{2}D_{b}\bar{\rho}\int_{{\bf k}}k^{4}U_{k}^{2}\left\langle e^{-i{\bf k}\cdot\left[{\bf x}(t)-{\bf x}(t')\right]}\right\rangle \nonumber \\
 & \times\frac{1}{\left(\tau_{b}/a_{k}\right)^{2}-1}\left[\tau_{b}e^{-\left|t-t'\right|/\tau_{b}}-a_{k}e^{-\left|t-t'\right|/a_{k}}\right]\label{eq:corr_eta_a}
\end{align}
It is observed that the noise correlations depend on the interactions
$U$(between bath and tracer) and $V$(between active bath particles).
In addition, they are also coupled with the tracer dynamics ${\bf x}\left(t\right)$. 

To proceed, we need to know the information of correlation $\left\langle e^{-i{\bf k}\cdot\left[{\bf x}(t)-{\bf x}(t')\right]}\right\rangle $.
In the short time limit, one may adopt the so-called ``adiabatic
approximation'' due to the time-scale separation between the tracer
motion and bath particle dynamics, such that the tracer motion is
not apparent and 
\begin{equation}
\left\langle e^{-i{\bf k}\cdot\left({\bf x}_{t}-{\bf x}_{t'}\right)}\right\rangle \approx1.\label{eq:AdApp}
\end{equation}
In this case, we have 
\begin{align}
 & \left\langle \boldsymbol{\eta}_{A}\left(t\right)\cdot\boldsymbol{\eta}_{A}\left(t'\right)\right\rangle \approx\mu_{t}^{2}D_{b}\bar{\rho}\int_{{\bf k}}\frac{k^{4}U_{k}^{2}}{\left(\tau_{b}/a_{k}\right)^{2}-1}\nonumber \\
 & \times\left[\tau_{b}e^{-\left|t-t'\right|/\tau_{b}}-a_{k}e^{-\left|t-t'\right|/a_{k}}\right]\label{eq:CorrA_Ad}
\end{align}
and 
\begin{equation}
\left\langle \boldsymbol{\eta}_{T}\left(t\right)\cdot\boldsymbol{\eta}_{T}\left(t'\right)\right\rangle \approx\mu_{t}^{2}\mu_{b}\bar{\rho}T\int_{{\bf k}}k^{4}U_{k}^{2}a_{k}e^{-\left|t-t'\right|/a_{k}}.\label{eq:CorrT_Ad}
\end{equation}
Note that the latter one is proportional to the memory kernel in the
GLE, Eq.(\ref{eq:GLE2}), i.e., 
\begin{equation}
\left\langle \boldsymbol{\eta}_{T}\left(t\right)\cdot\boldsymbol{\eta}_{T}\left(t'\right)\right\rangle =d\mu_{t}T\zeta\left(t-t'\right).
\end{equation}
Therefore, in the absence of activity (not considering the active
part $\boldsymbol{\eta}_{A}$), fluctuation-dissipation theorem(FDT)
holds among the systematic (memory) part and noisy part of the total
force exerted by the bath particles on the tracer, reminiscent of
the usual Brownian motion. Surely, in the presence of particle activity,
$\boldsymbol{\eta}_{A}$ takes effect and the conventional FDT does
not hold anymore since the system is far from equilibrium. 

Nevertheless, if long time dynamics is involved, such adiabatic approximation
is not valid. Noise correlations thus become quite complicated, depending
on the behavior of $\left\langle e^{-i{\bf k}\cdot\left[{\bf x}(t)-{\bf x}(t')\right]}\right\rangle $.
In the next section, we will mainly study the diffusion behavior of
the tracer particle. For that purpose, one may use ``Gaussian approximation'',
i.e., the tracer displacement within a time interval $t-t_{0}$ is
Gaussian distributed with variance $2D_{\text{eff}}\left(t-t_{0}\right)$,
such that 
\begin{equation}
\left\langle e^{-i{\bf k}\cdot\left[{\bf x}(t)-{\bf x}(t_{0})\right]}\right\rangle \approx e^{-\frac{k^{2}}{2}\left\langle \Delta{\bf x}^{2}(t)\right\rangle }=e^{-k^{2}dD_{{\rm eff}}\left(t-t_{0}\right)}\label{eq:GaussApp}
\end{equation}
wherein $\left\langle \Delta{\bf x}^{2}(t)\right\rangle =\left\langle \left[{\bf x}\left(t\right)-{\bf x}\left(0\right)\right]^{2}\right\rangle $
is the mean square displacement and $D_{\text{eff}}$ is the effective
(long time) diffusion coefficient of the tracer. Note however, $D_{\text{eff}}$
must be calculated from the tracer dynamics, which in turn depends
on the noise property and thus $D_{\text{eff}}$ itself. Therefore,
$D_{\text{eff}}$ should be calculated in a self-consistent way, which
will be discussed in the next section.

\section{Tracer Diffusion and Mobility}

In this section, we will investigate the long-time diffusion coefficient
$D_{\text{eff}}$ of the tracer. According to the Green-Kubo relation,
\begin{align}
D_{{\rm eff}}= & \lim_{t\rightarrow\infty}\frac{1}{2dt}\left\langle \Delta{\bf x}^{2}(t)\right\rangle =\frac{1}{d}\int_{0}^{\text{\ensuremath{\infty}}}{\rm d}t\left\langle \dot{{\bf x}}(t)\cdot\dot{{\bf x}}(0)\right\rangle \label{eq:Deff}
\end{align}
As already mentioned above, it is feasible to adopt the Markovian
approximation and the dynamics of ${\bf x}$ is governed by Eq.(\ref{eq:LE})
$\dot{{\bf x}}_{t}=\left(1+\lambda\right)^{-1}\left(\boldsymbol{\eta}_{A}+\boldsymbol{\eta}_{T}+\sqrt{2\mu_{t}T}\boldsymbol{\xi}_{t}\right)$.
Substitute this into Eq.(\ref{eq:Deff}), one may calculate $D_{\text{eff}}$
straightforwardly given the noise correlations of $\boldsymbol{\eta}_{A}$
and $\boldsymbol{\eta}_{T}$. Nevertheless, as shown in Eqs.(\ref{eq:corr_eta_t})
and (\ref{eq:corr_eta_a}), the noise correlations may depend on the
dynamics of ${\bf x}$ itself, thus proper approximations must be
employed. 

\subsection{Adiabatic Approximation}

If one simply uses the ``naive'' adiabatic approximation, Eq.(\ref{eq:AdApp}),
the subsequent noise correlations are given by Eqs.(\ref{eq:CorrA_Ad})
and (\ref{eq:CorrT_Ad}). One can thus obtain that 
\begin{align}
D_{\text{naive}} & =\frac{1}{d\left(1+\lambda\right)^{2}}\left(I_{1}^{\text{ad}}+I_{2}^{\text{ad}}+d\mu_{t}T\right)\nonumber \\
 & =\frac{1}{1+\lambda}\left(\mu_{t}T+\frac{\lambda}{1+\lambda}\mu_{T}\left(\frac{D_{b}}{\mu_{b}}\right)\right)\label{eq:Deff_0}
\end{align}
where 
\begin{align}
I_{1}^{\text{ad}}= & \int_{0}^{\infty}\left\langle \boldsymbol{\eta}_{A}\left(0\right)\cdot\boldsymbol{\eta}_{A}\left(t\right)\right\rangle {\rm d}t=\mu_{t}^{2}D_{b}\bar{\rho}\int_{{\bf k}}k^{4}U_{k}^{2}a_{k}\nonumber \\
= & d\mu_{t}\lambda\left(D_{b}/\mu_{b}\right)
\end{align}
and 
\begin{equation}
I_{2}^{\text{ad}}=\int_{0}^{\infty}\left\langle \boldsymbol{\eta}_{T}\left(0\right)\cdot\boldsymbol{\eta}_{T}\left(t\right)\right\rangle {\rm d}t=d\mu_{t}T\lambda,
\end{equation}
and use has been made $\int_{0}^{\infty}\left\langle \boldsymbol{\xi}_{T}\left(0\right)\cdot\boldsymbol{\xi}_{T}\left(t\right)\right\rangle {\rm d}t=d$.
The meaning of each term in Eq.(\ref{eq:Deff_0}) is clear. The first
term, $\mu_{t}T/\left(1+\lambda\right)$, gives the diffusivity of
the particle in a purely passive bath, wherein $\lambda$ accounts
for the excess friction that results from the interactions $U$ between
tracer and bath particles. The second term, $\left(D_{b}\mu_{t}/\mu_{b}\right)\lambda/\left(1+\lambda\right)^{2}$,
arises from the activity of the bath particles. One can see that $D_{b}/\mu_{b}$
acts as an effective temperature of the bath, which contributes an
active part to the total diffusion. However, as discussed above, this
approximation should only be valid for short time. Therefore, $D_{\text{naive}}$
only serves as the zero-order approximation of the long-time diffusion
constant. 

\subsection{Gaussian Approximation}

A better approximation would be the Gaussian one described in Eq.(\ref{eq:GaussApp}),
wherein $\left\langle e^{-i{\bf k}\cdot\left[{\bf x}(t)-{\bf x}(0)\right]}\right\rangle \simeq e^{-k^{2}dD_{{\rm eff}}t}$.
Using this approximation, one can obtain the noise correlations as
follows, 
\begin{multline}
\left\langle \boldsymbol{\eta}_{T}\left(0\right)\cdot\boldsymbol{\eta}_{T}\left(t\right)\right\rangle \approx\mu_{t}^{2}\mu_{b}\bar{\rho}T\int_{{\bf k}}k^{4}U_{k}^{2}a_{k}e^{-t\left(a_{k}^{-1}+k^{2}dD_{{\rm eff}}\right)}\label{eq:noise_gs_T}
\end{multline}
and
\begin{align}
\left\langle \boldsymbol{\eta}_{A}\left(0\right)\cdot\boldsymbol{\eta}_{A}\left(t\right)\right\rangle \approx & \mu_{t}^{2}D_{b}\bar{\rho}\int_{{\bf k}}\frac{k^{4}U_{k}^{2}e^{-k^{2}dD_{{\rm eff}}t}}{(\tau_{b}/a_{k})^{2}-1}\nonumber \\
 & \times\left[\tau_{b}e^{-t/\tau_{b}}-a_{k}e^{-t/a_{k}}\right].\label{eq:noise_gs_A}
\end{align}
Accordingly, the integral
\begin{align}
I_{1}^{\text{g}}= & \int_{0}^{\infty}\left\langle \boldsymbol{\eta}_{A}\left(0\right)\cdot\boldsymbol{\eta}_{A}\left(t\right)\right\rangle {\rm d}t\nonumber \\
= & \mu_{t}^{2}\bar{\rho}D_{b}\int_{{\bf k}}k^{4}U_{k}^{2}a_{k}^{2}F\left(k,D_{\text{eff}}\right),
\end{align}
with the factor
\begin{equation}
F\left(k,D_{\text{eff}}\right)=\frac{1+\frac{\tau_{b}a_{k}}{\tau_{b}+a_{k}}k^{2}dD_{\text{eff}}}{\left(1+a_{k}k^{2}dD_{{\rm eff}}\right)\left(1+\tau_{b}k^{2}dD_{{\rm eff}}\right)},\label{eq:Fk_Deff}
\end{equation}
and 
\begin{align}
I_{2}^{\text{g}} & =\int_{0}^{\infty}\left\langle \boldsymbol{\eta}_{T}\left(0\right)\cdot\boldsymbol{\eta}_{T}\left(t\right)\right\rangle {\rm d}t\nonumber \\
 & =\mu_{t}^{2}\bar{\rho}\left(\mu_{b}T\right)\int_{{\bf k}}k^{4}U_{k}^{2}a_{k}^{2}G\left(k,D_{\text{eff}}\right),
\end{align}
with 
\begin{equation}
G\left(k,D_{\text{eff}}\right)=\frac{1}{1+a_{k}k^{2}dD_{{\rm eff}}}.\label{eq:Gk_Deff}
\end{equation}

It is instructive to discuss more about these integrals of noise correlations.
Note that the time scale $a_{k}=\left[\mu_{b}k^{2}\left(T+\bar{\rho}V_{k}\right)\right]^{-1}$
characterizes the $k$-mode density fluctuation of the bath that exerts
on the tracer particle, which decays fast ($\sim k^{-2}$) with increasing
$k$. On the other hand, $\tau_{b}$ gives the persistence time of
the bath particle, which results from the active feature of the bath.
Within $\tau_{b}$, an isolated bath particle will move a persistence
length given approximately by $\sqrt{D_{b}\tau_{b}}$. Identifying
$\tau_{b}=a_{k_{b}}$ gives us a characteristic wave factor $k_{b}$,
which determines the length scale ($k_{b}^{-1}$) that approximately
matches the persistence length. 

One may then divide the range of $k$ into three ranges: (i) $k\ll k_{b}$,
(ii) $k\sim k_{b}$ and (iii) $k\gg k_{b}$, where $F\left(k\right)/G\left(k\right)$
behaves differetly. In range (i) for very small $k$ (long wave-length)
, $a_{k}\gg\tau_{b}$, such that $F\left(k\right)/G\left(k\right)=1$.
In this range, $I_{1}^{g}/I_{2}^{g}=D_{b}/\left(\mu_{b}T\right)$,
i.e, the effect of acvitity is reflected by an effective temperature
$D_{b}/\mu_{b}$, same as that in the adiabatic approximation. In
range (iii) for large $k$, $a_{k}\ll\tau_{b}$ and $F\left(k\right)/G\left(k\right)=\left(1+\tau_{b}k^{2}dD_{\text{eff}}\right)^{-1}\ll1$.
Note that in this range, both integrals $I_{1}^{g}$ and $I_{2}^{g}$
are very small and can be neglected($U_{k\to\infty}\to0$). Some complexity
arises in range (ii), wherein $a_{k}\sim\tau_{b}$ and $F\left(k\right)/G\left(k\right)\simeq\left(1+\frac{1}{2}k^{2}dD_{\text{eff}}\right)/\left(1+\frac{1}{2}k^{2}dD_{\text{eff}}\right)\sim\frac{1}{2}$.
In this moderate range of $k$, $I_{1}^{g}/I_{2}^{g}\simeq D_{b}/2\mu_{b}T$. 

\subsection{Self-Consistent Equatoin for Diffusivity}

Finally, we can obtain a self-consistent equation for the effective
diffusion constant 
\begin{align}
\left(1+\lambda\right)^{2}D_{\text{eff}}= & \frac{1}{d}\left(I_{\text{1}}^{g}+I_{2}^{g}+d\mu_{t}T\right)\label{eq:Deff_sc}\\
= & \frac{\mu_{t}^{2}\bar{\rho}}{d}\int_{{\bf k}}k^{4}U_{k}^{2}a_{k}^{2}\Big[D_{b}F\left(k,D_{\text{eff}}\right)\nonumber \\
 & +\mu_{b}TG\left(k,D_{\text{eff}}\right)\Big]+\mu_{t}T\nonumber 
\end{align}
which is the second main result of the present work, serves as a starting
point for numerical calculation of $D_{\text{eff}}$. If one simply
sets $F\left(k,D_{\text{eff}}\right)\simeq G\left(k,D_{\text{eff}}\right)\simeq1$,
it recovers the result from adiabatic approximation, Eq.(\ref{eq:Deff_0}).
As discussed in the last paragraph, generally $F\left(k,D_{\text{eff}}\right)\le G\text{\ensuremath{\left(k,D_{\text{eff}}\right)}}<1$,
thus $D_{\text{eff}}$ will be smaller than the naive value $D_{\text{naive}}$. 

Compare $I_{1,2}^{g}$ with $I_{1,2}^{\text{ad}}$, one can see that
the self-consistent correction amounts to the factors $F\left(k,D_{\text{eff}}\right)$
or $G\left(k,D_{\text{eff}}\right)$. Generally, such a correction
becomes more apparent if $D_{\text{eff}}$ is larger. Roughly, if
$D_{b}$ is very small, $I_{1}^{g}\simeq I_{1}^{\text{ad}}\propto D_{b}$,
such that $D_{\text{eff}}$ increases linearly with $D_{b}$. Nevertheless,
for large $D_{b}$, $I_{2}^{g}$(or $I_{2}^{\text{ad}}$) would be
approximately proportional to $1/D_{\text{eff}}$. As a consequence,
$D_{\text{eff}}\propto\left(D_{b}+c\right)/D_{\text{eff}}$ ($c$
is a parameter not dependent on $D_{b}$), indicating that $D_{\text{eff}}$
scales approximately as $\sqrt{D_{b}}$. If mapping to an active Brownian
particle(ABP) model, $D_{b}$ here would be proportional to $v_{0}^{2}$
(or ${\rm Pe}^{2}$) with $v_{0}$ the self-propulsion velocity of
ABP and ${\rm Pe}$ the Péclet number. Therefore, our theory predicts
that $D_{\text{eff}}\propto{\rm Pe}^{2}$ for small ${\rm Pe}$ and
$D_{\text{eff}}\propto{\rm Pe}$ for large ${\rm Pe}$. In Burkholder
and Brady's work\citep{2017_PRE_TracerDiffActBath}, they studied
the tracer diffusion in a dilute dispersion of active particles by
using Smoluchowski-level analysis. They also found that the diffusivity
scales as ${\rm Pe^{2}}$ when activity is weak and as ${\rm Pe}$
when activity is strong, consistent with our result here. 

On the relationship between effective diffusion coefficient $D_{{\rm eff}}$
and bath density $\bar{\rho}$, in the early experimental works, Wu
et al\citep{2000_PRL_Wu_bacteria_bath} found that the tracer diffusion
in bacterial bath increases linearly with the bacterial density at
dilute region, which is also supported by the works of Kim \citep{2004_PF_Kim_EnhancedDiff},
Lepto and coworkers\citep{2009_PRL_Goldstein_EnhancedTrcDiff}.In
our model, as shown in Eq.\eqref{eq:Deff_0}, $D_{{\rm naive}}$ increases
with $\bar{\rho}$ firstly and then decreases, due to the $\frac{\lambda}{\left(1+\lambda\right)^{2}}$
term and $\lambda\propto\bar{\rho}$. Therefore, the leading order
of effective diffusion indeed increases with $\bar{\rho}$ when it
is very small, in consistent with previous researches. It can also
be shown easily that $D_{{\rm eff}}\propto\bar{\rho}$ when $\bar{\rho}$
is small. 

\subsection{Effective Mobility}

Another important property of the tracer is the mobility, characterizing
the response of the tracer velocity to external forces. Note that
in equilibrium systems, mobility $\mu$ and diffusion coefficient
$D$ are related to each other via the Einstein relation, $D=\mu k_{B}T$,
which is also known as fluctuation-dissipation theorem for a Brownian
particle. Nevertheless, for nonequilibrium systems, such a FDT may
not hold. In fact, there have been some recent studies about generalized
linear-response or FDT in nonequilibrium active systems. However,
the mobility of the tracer has not been calculated explicitly so far. 

Here we would like to calculate the tracer mobility by using the GLE,
Eq.\eqref{eq:GLE1}. According to the definition, we apply a small
external constant force ${\bf f}$ on the tracer and then calculate
the averaged velocity $\left\langle \dot{x}^{\shortparallel}\right\rangle $
(without external force, $\left\langle \dot{x}^{\shortparallel}\right\rangle $
would be zero) along the force direction(denoted by '$\shortparallel$').
The mobility $\mu_{{\rm eff}}$ is then obtained as
\begin{equation}
\mu_{{\rm eff}}=\lim_{f\rightarrow0}\frac{\left\langle \dot{x}_{t}^{\shortparallel}\right\rangle }{\text{\ensuremath{\left|{\bf f}\right|}}}\label{eq:mu_eff}
\end{equation}
Adding a dragging force term $\mu_{t}{\bf f}$ to the right-hand side
of Eq.(\eqref{eq:GLE1}) and then taking the long time average (the
noise terms all vanish), we get the equation for averaged velocity
at the direction of ${\bf f}$ as
\begin{align}
\left\langle \dot{x}_{t}^{\shortparallel}\right\rangle = & \mu_{t}f-\mu_{t}\int_{-\infty}^{t}\mu_{b}\bar{\rho}\int_{{\bf k}}ik_{\shortparallel}k^{2}U_{k}^{2}\nonumber \\
 & \times\left\langle e^{-i{\bf k}\cdot\left[{\bf x}\left(t\right)-{\bf x}\left(s\right)\right]}\right\rangle e^{-\left(t-s\right)/a_{k}}{\rm d}s\label{eq:velo}
\end{align}
Again, one needs to make approximations to the term $\left\langle e^{-i{\bf k}\cdot\left[{\bf x}\left(t\right)-{\bf x}\left(s\right)\right]}\right\rangle $.
In the above sections, we have argued that the Gaussian approximation,
i.e, $\left\langle e^{-i{\bf k}\cdot\left[{\bf x}\left(t\right)-{\bf x}\left(s\right)\right]}\right\rangle \simeq e^{-k^{2}dD_{{\rm eff}}\left|t-s\right|}$,
is appropriate for long time dynamics. Here we may use the same spirit
for mobility calculation. Note however, the deterministic motion caused
by the dragging force ${\bf f}$ must be taken out. Therefore, we
apply the Gaussian approximation to the remaining dynamics apart from
the dragging motion(for $t>s$), i.e.,
\begin{equation}
\left\langle e^{-i{\bf k}\cdot\left[{\bf x}\left(t\right)-{\bf x}\left(s\right)\right]}\right\rangle \simeq e^{-ik_{\shortparallel}\mu_{f}f\left(t-s\right)}e^{-k^{2}dD_{{\rm eff}}(t-s)}
\end{equation}
wherein $\mu_{f}$ denotes the $f$-dependent mobility, which is related
to the target $\mu_{\text{eff}}$ via $\mu_{\text{eff}}=\lim_{f\to0}\mu_{f}$.
Substituting this correlation back into Eq.(\ref{eq:velo}) and integrating
over time, we can get a self-consistent equation for $\mu_{f}$ as
,
\begin{equation}
\mu_{f}=\mu_{t}-\mu_{t}\mu_{b}\bar{\rho}\int_{{\bf k}}\frac{k_{\shortparallel}^{2}\mu_{f}k^{2}U_{k}^{2}}{k^{4}\left[\mu_{b}\left(T+\bar{\rho}V_{k}\right)+dD_{{\rm eff}}\right]^{2}+k_{\shortparallel}^{2}\mu_{f}^{2}f^{2}}
\end{equation}
Then taking the limit $f\rightarrow0$, we obtain a self-consistent
equation for the effective mobility (using $k^{2}=dk_{\shortparallel}^{2}$)
\begin{equation}
\mu_{{\rm eff}}=\mu_{t}-\frac{\mu_{t}\mu_{b}\bar{\rho}}{d}\int_{{\bf k}}\frac{U_{k}^{2}\mu_{{\rm eff}}}{\left[\mu_{b}\left(T+\bar{\rho}V_{k}\right)+dD_{{\rm eff}}\right]^{2}}\label{eq:mu_eff_sc}
\end{equation}
which is the third main result of the present work. Notice that we
have to calculate $D_{\text{eff}}$ first to calculate $\mu_{\text{eff}}$. 

With both $D_{\text{eff}}$ and $\mu_{\text{eff}}$ calculated, one
may thus define an effective temperature through Stokes-Einstein-Sutherland
relation, i.e., 
\begin{equation}
T_{\text{eff}}=D_{\text{eff}}/\mu_{\text{eff}}\label{eq:Teff}
\end{equation}
It is interesting to investigate how such a defined effective temperature
depends on the bath properties. 

\subsection{Numerical Simulation}

In above sections, we have derived self-consistent equations for the
effective (long time) diffusion constant and mobility of the tracer
particle, which can be easily calculated numerically. Herein, we would
like to compare the numerical results with direct simulations of the
original system. The simulations proceed in a two dimensional square
box with length $L$ and periodic boundaries. We set the size of bath
particles $R_{b}$ as the unit of length, $T$ as the unit of energy,
and $R_{b}^{2}/(\mu_{b}T)$ as the unit of time. The interaction between
particles are given by the soft harmonic potential, i.e., $U(r)=\epsilon_{U}|r-\sigma_{tb}|^{2}$
and $V(r)=\epsilon_{V}|r-\sigma_{bb}|^{2}$, where $\epsilon_{U}$($\epsilon_{V}$)
denotes potential strength, $\sigma_{bb}=R_{b},$ and $\sigma_{tb}=(R_{tr}+R_{b})/2$
with $R_{tr}$ the tracer radius. In simulations, tracer diffusivity
is calculated with standard method, i.e. time derivative of the mean
square displacement in the long time limit. For mobility, we apply
a small force to the tracer and calculate the velocity change directly
according to the definition.  

\begin{figure}
(a)\includegraphics[width=7cm]{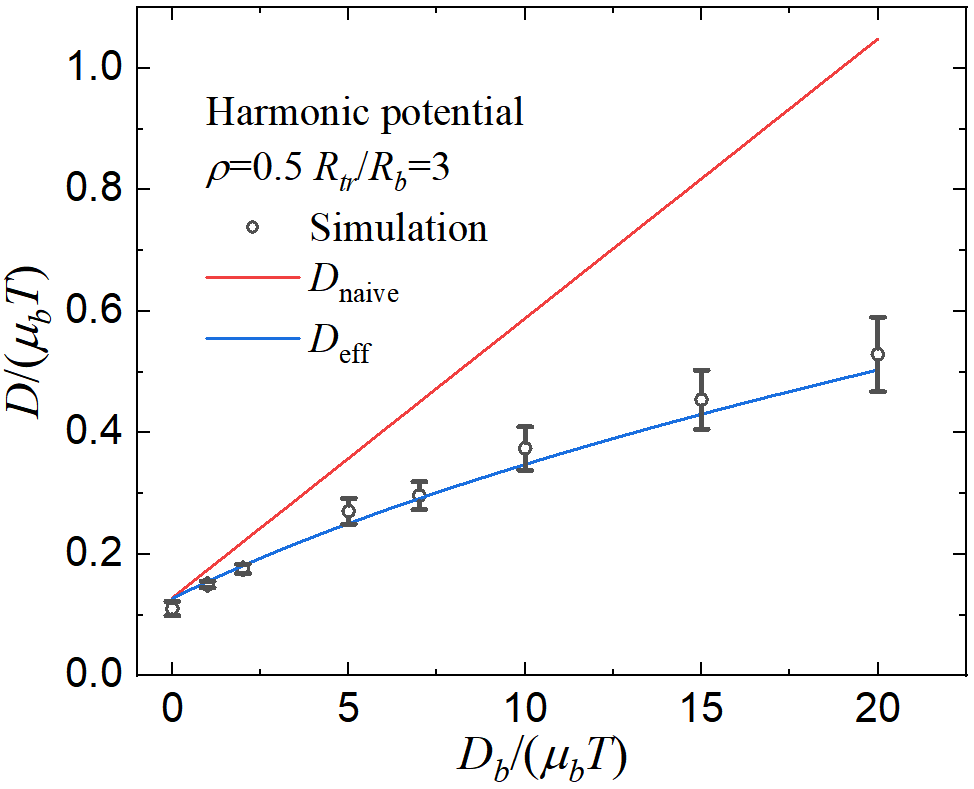}

(b)\includegraphics[width=7cm]{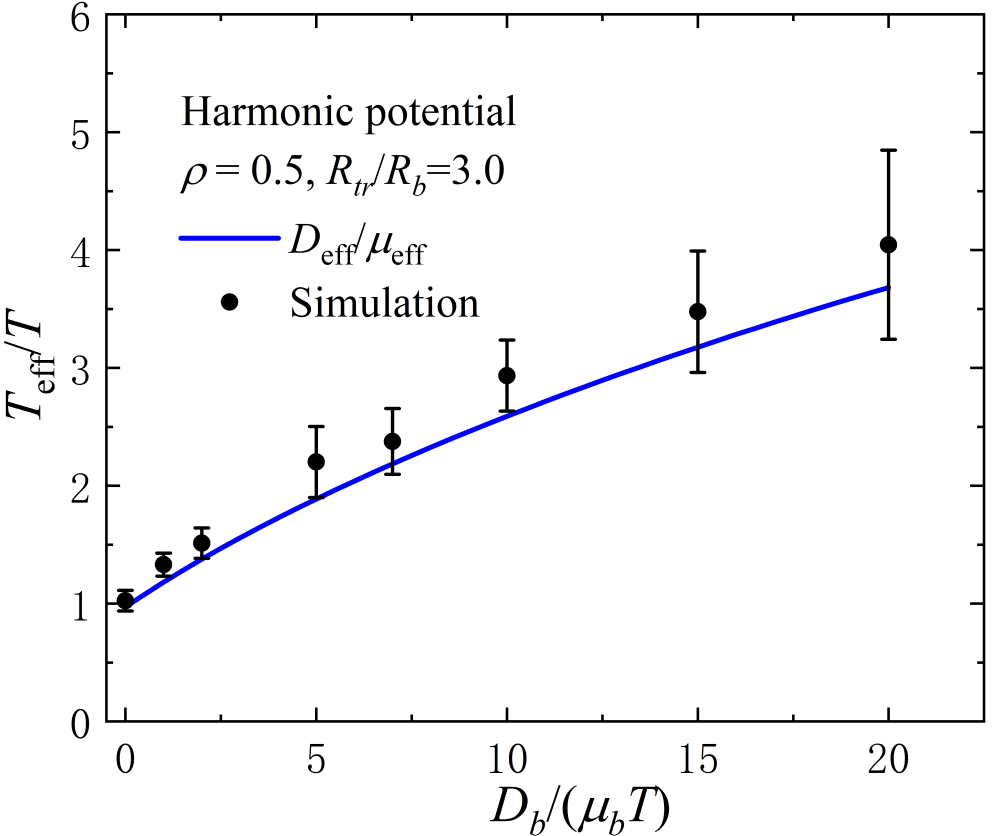}

\caption{Effective diffusion coefficient with bath activity for different conditions:
(a)harmonic potentials and number density of active bath is $\rho=0.5$
, red line shows the linearly dependent of $D_{{\rm naive}}$ on
$D_{b}$, and blue line is the effective diffusivity $D_{{\rm eff}}$
with the self-consistent correction. Round circles are simulation
data in the same parameters. (b) Effective temperature defined as
$T_{{\rm eff}}=D_{{\rm eff}}/\mu_{{\rm eff}}$, the parameters are
the same as (a). Simulation details: in this case, we choose $L=50R_{b}$,
$\epsilon_{U}:\epsilon_{V}:T=20:100:1$, $\mu_{tr}/\mu_{b}=0.2$,
$\tau_{p}=0.5R_{b}^{2}/(\mu_{b}T)$.}
\label{fig:Deff}
\end{figure}

In Fig.\ref{fig:Deff}(a), the dependence of tracer diffusivity is
shown as a function of bath activity $D_{b}$, for typical parameter
settings. Symbols with error bars are obtained from direct simulations,
and the blue line is obtained from Eq.(\ref{eq:Deff_sc}). Clearly,
our theory is in good agreement with simulations. If instead, one
just use the naive approximation for the diffusivity, which is given
by the red line for $D_{\text{naive}}$, the discrepancy becomes larger
and larger with increasing $D_{b}$. Therefore, our self-consistent
equation for $D_{\text{eff}}$ is necessary to account for the nonlinear
feedback effects of bath activity. As discussed above, $D_{\text{eff}}$
scales linearly with $D_{b}$ for small $D_{b}$, while $D_{\text{eff}}\sim D_{b}^{1/2}$
for large $D_{b}$, which is evidently shown in the figure. In \ref{fig:Deff}(b),
data for the effective temperature $T_{\text{eff}}=D_{\text{eff}}/\mu_{\text{eff}}$
is also shown as a function of the bath activity $D_{b}$. Again,
good agreements between the simulation (symbols) and numerical (line)
results are also evident. We have also performed simulations with
other parameter settings and other soft potential such as Gaussian
potential, and good quantitative agreements between our self-consistent
equations, (\ref{eq:Deff_sc}) and (\ref{eq:Teff}), and simulation
results are also observed, while numerical calculations are much much
faster than direct simulations. 

\section{Conclusion}

Motivated by a series of works about tracer in nonequilibrium baths,
we presented a theoretical method to investigate the transport properties
of the tracer in such environments, including effective diffusion
and mobility. The theory begins with the equations of motion for tracer
and driven bath particles. Using the procedure of linearized Dean's
equation and the mean-field theory approximation, we derived a generalized
Langevin equation for the tracer, wherein a memory kernel and two
nontrivial colored noises are involved.  To understand the relationships
of mobility and diffusivity with the properties of the driven bath,
we firstly introduced an adiabatic approximation to calculate these
transport coefficients. However, these results are only valid when
time-scales of tracer and bath particles are clearly separated. For
more general situations, we considered the associated long-time behavior
and introduced the Gaussian approximation. As a result, self-consistent
equations for effective diffusion and mobility are derived. Numerical
calculation shows good conformance with simulation data. We also defined
an effective temperature through the quotient of diffusivity and mobility,
the theoretical result also agrees with the simulation. 

For future works, we wish to study more statistical properties of
the system such as linear response and fluctuation theorem. For more
in-depth research, we also want to study the relationship between
transport properties and {\small{}the }bath properties, such as density,
persistent time of particle, or other kinds of active particles. Further
we want to reform the system to a Brownian heat engine in active bath.
We also notice that the non-Gaussian distribution behavior of a tracer
in active baths \citep{2009_PRL_Goldstein_EnhancedTrcDiff,2017_SciRep_Maggi_MemLessResponseAndFDT},
and hope our theory can bring theoretical understanding about this
issue. We may also use this theoretical framework to study multi tracers
system and investigate the effective interactions between tracers.

This work is supported by MOST(2018YFA0208702), NSFC (32090044, 21973085,
21833007, 21790350, 21521001), Anhui Initiative in Quantum Information
Technologies (AHY090200), and the Fundamental Research Funds for the
Central Universities (WK2340000104).

\bibliographystyle{unsrt}
\bibliography{ActiveBath,ActiveGeneral,DiffFDT}

\onecolumngrid

\appendix

\section{Derivation Details\label{sec:detail}}

\subsection{Derivation of Linearized Dean's equation\label{subsec:Deri_dean}}

Firstly, we address the one-dimensional model to briefly illustrate
the derivation of Eq.\eqref{eq:dean}. The generalization to higher
dimensional system is trivial. Consider a stochastic differential
equation(SDE), $\frac{{\rm d}x}{{\rm d}t}=a(x,t)+b(x,t)\eta(t)$ where
$\left\langle \eta(t)\right\rangle =0$ and $\left\langle \eta(t)\eta(t')\right\rangle =\delta(t-t')$
denote the Gaussian white noise. Using the It${\rm \hat{o}}$ calculus,
for any well behaved function $f(x)$, one has
\begin{equation}
\frac{{\rm d}f(x)}{{\rm d}t}=\left[a(x,t)\partial_{x}f+\frac{b^{2}(x,t)}{2}\partial_{x}^{2}f\right]+\left(\partial_{x}f\right)b(x,t)\eta(t)
\end{equation}
Now substituting the Langevin equation \eqref{eq:LE_Bath} of bath
particles for the SDE above, using the It${\rm \hat{o}}$ calculus,
we have an evolution equation for arbitrary function $g({\bf r}_{i})$
\begin{align}
\frac{{\rm d}g\left({\bf r}_{i}\right)}{{\rm d}t}= & \sqrt{2\mu_{b}T}\boldsymbol{\xi}_{i}\cdot\nabla_{i}g+\left(\nabla_{i}g\right)\left[-\mu_{b}\nabla_{i}\left(\sum_{j\neq i}V+U\right)+{\bf f}_{i}\right]+\mu_{b}T\nabla_{i}^{2}g\nonumber \\
= & \int{\rm d}{\bf r}\rho_{i}({\bf r},t)\left\{ \sqrt{2\mu_{b}T}\boldsymbol{\xi}_{i}\cdot\nabla g+\left(\nabla g\right)\left[-\mu_{b}\nabla\left(\int{\rm d}{\bf r}'\rho({\bf r}',t)V\left(\left|{\bf r}-{\bf r}'\right|\right)+U\left(\left|{\bf r}-{\bf x}\right|\right)\right)+{\bf f}_{i}\right]+\mu_{b}T\nabla^{2}g\right\} \nonumber \\
= & \int{\rm d}{\bf r}g({\bf r})\left\{ -\sqrt{2\mu_{b}T}\nabla\rho_{i}\cdot\boldsymbol{\xi}_{i}-\nabla\cdot\left[{\bf f}_{i}\rho_{i}({\bf r},t)\right]+\mu_{b}\nabla\rho_{i}({\bf r},t)\right.\nonumber \\
 & \cdot\left.\left[\nabla\left(\int{\rm d}{\bf r}'\rho({\bf r}',t)V\left(\left|{\bf r}-{\bf r}'\right|\right)+U\left(\left|{\bf r}-{\bf x}\right|\right)\right)\right]+\mu_{b}T\nabla^{2}\rho_{i}({\bf r},t)\right\} 
\end{align}
wherein the bath density $\rho\left({\bf r},t\right)=\sum_{i}\rho_{i}\left({\bf r},t\right)=\sum_{i}\delta\left({\bf r}-{\bf r}_{i}\left(t\right)\right)$
is applied. On the other hand, we can also write $\frac{{\rm d}g({\bf r}_{i})}{{\rm d}t}=\int\frac{\partial\rho_{i}({\bf r},t)}{\partial t}g({\bf r}){\rm d}{\bf r}$.
Considering the arbitrariness of function $g({\bf r})$, we have 
\begin{equation}
\frac{\partial\rho_{i}\left({\bf r},t\right)}{\partial t}=-\sqrt{2\mu_{b}T}\nabla\rho_{i}\cdot\boldsymbol{\xi}_{i}-\nabla\cdot\left[{\bf f}_{i}\rho_{i}\right]+\mu_{b}\nabla\rho_{i}\cdot\left[\nabla\left(\int{\rm d}{\bf r}'\rho({\bf r}',t)V\left(\left|{\bf r}-{\bf r}'\right|\right)+U\left(\left|{\bf r}-{\bf x}\right|\right)\right)\right]+\mu_{b}T\nabla^{2}\rho_{i}
\end{equation}
and the collective density 
\begin{equation}
\frac{\partial\rho\left({\bf r},t\right)}{\partial t}=-\sum_{i}\sqrt{2\mu_{b}T}\nabla\cdot\left[\boldsymbol{\xi}_{i}\rho_{i}\right]-\sum_{i}\nabla\cdot\left[{\bf f}_{i}\rho_{i}\right]+\mu_{b}\nabla\rho\cdot\left[\nabla\left(\int{\rm d}{\bf r}'\rho({\bf r}',t)V\left(\left|{\bf r}-{\bf r}'\right|\right)+U\left(\left|{\bf r}-{\bf x}\right|\right)\right)\right]+\mu_{b}T\nabla^{2}\rho\label{eq:den_not_cns}
\end{equation}
notice that this equation is not closed for the moment, due to the
first two noise terms. To achieve Eq.\eqref{eq:Dean}, we need to
introduce the noise field to replace these terms, meanwhile keeping
the correlation properties invariant. Let $\chi_{1}(t)=-\sum_{i}\sqrt{2\mu_{b}T}\nabla\cdot\left[\boldsymbol{\xi}_{i}(t)\rho_{i}({\bf r},t)\right]$,
$\chi_{2}(t)=-\sum_{i}\nabla\cdot\left[{\bf f}_{i}(t)\rho_{i}({\bf r},t)\right]$,
the time correlation function of these terms are
\begin{align*}
\left\langle \chi_{1}(t)\chi_{1}(t')\right\rangle = & 2\mu_{b}T\delta(t-t')\sum_{i=1}^{N}\nabla_{r}\cdot\nabla_{r'}\left[\rho_{i}({\bf r},t)\rho_{i}({\bf r}',t')\right]\\
= & 2\mu_{b}T\delta(t-t')\nabla_{r}\cdot\nabla_{r'}\left[\sum_{i=1}^{N}\rho_{i}({\bf r},t)\delta({\bf r}-{\bf r}')\right]\\
= & 2\mu_{b}T\delta(t-t')\nabla_{r}\cdot\nabla_{r'}\left[\rho\left({\bf r},t\right)\delta\left({\bf r}-{\bf r}'\right)\right]
\end{align*}
\begin{align*}
\left\langle \chi_{2}(t)\chi_{2}(t')\right\rangle = & \frac{D_{b}}{\tau_{b}}e^{-\left|t-t'\right|/\tau_{b}}\sum_{i=1}^{N}\nabla_{r}\cdot\nabla_{r'}\left[\rho_{i}({\bf r},t)\rho_{i}({\bf r}',t')\right]\\
= & \frac{D_{b}}{\tau_{b}}e^{-\left|t-t'\right|/\tau_{b}}\nabla_{r}\cdot\nabla_{r'}\left[\sum_{i=1}^{N}\rho_{i}({\bf r},t)\delta({\bf r}-{\bf r}')\right]\\
= & \frac{D_{b}}{\tau_{b}}e^{-\left|t-t'\right|/\tau_{b}}\nabla_{r}\cdot\nabla_{r'}\left[\rho\left({\bf r},t\right)\delta\left({\bf r}-{\bf r}'\right)\right]
\end{align*}
Now we introduce two global noise fields, $\chi_{1}'\left({\bf r},t\right)=\nabla\cdot\left[\sqrt{\rho\left({\bf r},t\right)}\boldsymbol{\xi}^{T}\right]$
and $\chi_{2}'\left({\bf r},t\right)=\nabla\cdot\left[\sqrt{\rho\left({\bf r},t\right)}\boldsymbol{\eta}_{f}\right]$,
where the correlations are $\left\langle \boldsymbol{\xi}^{T}\left({\bf r},t\right)\boldsymbol{\xi}^{T}\left({\bf r}',t'\right)\right\rangle =2\mu_{b}T\delta\left(t-t'\right)\delta\left({\bf r}-{\bf r}'\right){\bf 1}$
and $\left\langle \boldsymbol{\xi}^{A}\left({\bf r},t\right)\boldsymbol{\xi}^{A}\left({\bf r}',t'\right)\right\rangle =\frac{D_{b}}{\tau_{b}}\delta\left({\bf r}-{\bf r}'\right){\bf 1}$
respectively. It is easy to find out the $\chi_{1}$and $\chi_{1}'$
have the same time correlation function, also true for $\chi_{2}$
and $\chi_{2}'$. Now we find the replacement and Eq.\eqref{eq:den_not_cns}
becomes 
\begin{equation}
\frac{\partial\rho\left({\bf r},t\right)}{\partial t}=\nabla\cdot\left[\sqrt{\rho\left({\bf r},t\right)}\boldsymbol{\xi}^{T}\right]+\nabla\cdot\left[\sqrt{\rho\left({\bf r},t\right)}\boldsymbol{\xi}^{A}\right]+\mu_{b}\nabla\rho\cdot\left[\nabla\left(\int{\rm d}{\bf r}'\rho({\bf r}',t)V\left(\left|{\bf r}-{\bf r}'\right|\right)+U\left(\left|{\bf r}-{\bf x}\right|\right)\right)\right]+\mu_{b}T\nabla^{2}\rho\label{eq:dean}
\end{equation}
i.e. Eq.\eqref{eq:Dean}, which is a self-consistent equation. 

Although we use this mean-field approximation to simply the time evolution
equation of collective density, this equation is still to hard to
solve. For a system that the density fluctuation is small, we can
further assume that $\sqrt{\rho\left({\bf r},t\right)}\approx\sqrt{\bar{\rho}}$
where $\bar{\rho}=N/V$ is the averaged number density. This approximation
is also used to simplify the interaction terms, $\rho({\bf r})\int{\rm d}{\bf r}'\rho({\bf r}')\nabla V\left(\left|{\bf r}-{\bf r}'\right|\right)\approx\bar{\rho}\rho({\bf r})*\nabla_{r}V({\bf r})$
and $\rho({\bf r})\nabla U\left(\left|{\bf r}-{\bf x}\right|\right)\approx\bar{\rho}\nabla U\left(\left|{\bf r}-{\bf x}\right|\right)$,
which is $i{\bf k}e^{i{\bf k}\cdot{\bf x}}\bar{\rho}U_{k}$ in Fourier
space. Using the Fourier transform $\delta\rho_{k}(t)=\int e^{i{\bf k}\cdot{\bf r}}\left[\rho\left({\bf r},t\right)-\bar{\rho}\right]{\rm d}{\bf r}$,
we have 
\begin{equation}
\frac{\partial\delta\rho_{k}\left(t\right)}{\partial t}=-\mu_{b}k^{2}\left[\left(T+\bar{\rho}V_{k}\right)\delta\rho_{k}\left(t\right)+\bar{\rho}U_{k}e^{i{\bf k}\cdot{\bf x}}\right]+\sqrt{\bar{\rho}}i{\bf k}\cdot\left[\tilde{\boldsymbol{\xi}}_{k}^{T}\left(t\right)+\tilde{\boldsymbol{\xi}}_{k}^{A}\left(t\right)\right]
\end{equation}
where $V_{k}$ , $U_{k}$, $\tilde{\boldsymbol{\xi}}_{k}^{T}$ and
$\tilde{\boldsymbol{\xi}}_{k}^{A}$ are Fourier transforms of $V(r)$,
$U(r)$, $\boldsymbol{\xi}^{T}$ and $\boldsymbol{\xi}^{A}$ respectively.
This is the Eq.\eqref{eq:dean_flu} in the main text. 

\subsection{Derivation of generalized Langevin equation \label{subsec:gle}}

From equation \eqref{eq:GLE1},
\begin{align}
 & \mu_{t}\int_{-\infty}^{t}{\bf F}\left({\bf x}\left(t\right)-{\bf x}\left(s\right),t-s\right){\rm d}s\nonumber \\
= & -\mu_{t}\mu_{b}\bar{\rho}\int_{-\infty}^{t}{\rm d}s\int_{{\bf k}}i{\bf k}k^{2}U_{k}^{2}e^{-i{\bf k}\cdot\left[{\bf x}\left(t\right)-{\bf x}(s)\right]}e^{-\left(t-s\right)/a_{k}}\nonumber \\
\approx & -\frac{\mu_{t}\mu_{b}\bar{\rho}}{d}\int_{-\infty}^{t}{\rm d}s\int_{{\bf k}}k^{4}U_{k}^{2}e^{-\left(t-s\right)/a_{k}}\left[\int_{s}^{t}{\rm d}u\dot{{\bf x}}(u)\right]\nonumber \\
= & -\frac{\mu_{t}\mu_{b}\bar{\rho}}{d}\int_{-\infty}^{t}{\rm d}u\int_{-\infty}^{u}{\rm d}s\dot{{\bf x}}(u)\int_{{\bf k}}k^{4}U_{k}^{2}e^{-\left(t-s\right)/a_{k}}\nonumber \\
= & -\frac{\mu_{t}\mu_{b}\bar{\rho}}{d}\int_{-\infty}^{t}{\rm d}u\dot{{\bf x}}(u)\int_{{\bf k}}k^{4}U_{k}^{2}a_{k}e^{-\left(t-u\right)/a_{k}}
\end{align}
The approximation is about the term $e^{-i{\bf k}\cdot\left[{\bf x}\left(t\right)-{\bf x}(s)\right]}$.
According to the symmetry properties of $U_{k}$ and $a_{k}$ in ${\bf k}$-space,
only odd terms of $k$ in $e^{-i{\bf k}\cdot\left[{\bf x}\left(t\right)-{\bf x}(s)\right]}$
contribute to the integral. Using the Taylor expansion and omitting
the $O(k^{3})$ terms, we write $e^{-i{\bf k}\cdot\left[{\bf x}\left(t\right)-{\bf x}(s)\right]}\rightarrow-i{\bf k}\cdot\left[{\bf x}\left(t\right)-{\bf x}(s)\right]=-i{\bf k}\cdot\int_{s}^{t}{\rm d}u\dot{{\bf x}}(u)$.
With this approximation, and then exchange the order of integrals
over $u$ and $s$, we get the finial result.

\subsection{Derivation of correlation function of colored noise\label{subsec:actnoi}}

Firstly we emphasis that the size of the tracer particle is apparently
large than that of the bath particles. This leads to a shorter characteristic
time scale for the evolution of the tracer than the bath particles
as well. This understanding helps us in the following derivation.
The correlations of noise $\boldsymbol{\eta}_{A,T}(t)$ is written
as
\begin{align}
 & \left\langle \boldsymbol{\eta}_{A,T}\left(t\right)\cdot\boldsymbol{\eta}_{A,T}\left(t'\right)\right\rangle \nonumber \\
= & \mu_{t}^{2}\bar{\rho}\int_{{\bf k}}\int_{{\bf q}}{\bf k}\cdot{\bf q}U_{k}U_{q}\int_{-\infty}^{t}{\rm d}s\int_{-\infty}^{t'}{\rm d}s'e^{-\left(t-s\right)/a_{k}-\left(t'-s'\right)/a_{q}}\left\langle e^{-i{\bf k}\cdot{\bf x}(t)-i{\bf q}\cdot{\bf x}(t')}{\bf k}\cdot\tilde{\boldsymbol{\xi}}_{k}^{A,T}(s){\bf q}\cdot\tilde{\boldsymbol{\xi}}_{q}^{A,T}(s')\right\rangle \nonumber \\
\approx & \mu_{t}^{2}\bar{\rho}\int_{{\bf k}}\int_{{\bf q}}{\bf k}\cdot{\bf q}U_{k}U_{q}\int_{-\infty}^{t}{\rm d}s\int_{-\infty}^{t'}{\rm d}s'e^{-\left(t-s\right)/a_{k}-\left(t'-s'\right)/a_{q}}\left\langle e^{-i{\bf k}\cdot{\bf x}(t)-i{\bf q}\cdot{\bf x}(t')}\right\rangle \left\langle {\bf k}\cdot\tilde{\boldsymbol{\xi}}_{k}^{A,T}(s){\bf q}\cdot\tilde{\boldsymbol{\xi}}_{q}^{A,T}(s')\right\rangle \label{eq:corrAT}
\end{align}
Herein, the reason why the correlation can be divided into two parts
is that ${\bf x}(t)$ can be seen as a slow variable and $\tilde{\boldsymbol{\xi}}_{k}^{A,T}(s)$
are fast variables, i.e. there is a time scale separation between
${\bf x}(t)$ and $\tilde{\boldsymbol{\xi}}_{k}^{A,T}(s)$. At the
typical time scale of ${\bf x}(t)$ evolution, noise $\tilde{\boldsymbol{\xi}}_{k}^{A,T}(s)$
has been sufficiently averaged, which is zero. Finally we have the
form above.

Notice that the Fourier transform of a white noise herein is actually
firstly calculating the correlation and then proceeding the transform.
Therefore,
\begin{align}
 & \left\langle {\bf k}\cdot\tilde{\boldsymbol{\xi}}_{k}^{T}(s){\bf q}\cdot\tilde{\boldsymbol{\xi}}_{q}^{T}(s')\right\rangle \nonumber \\
= & {\bf k}\cdot\iint{\rm d}{\bf r}{\rm d}{\bf r}'e^{i{\bf k}\cdot{\bf r}}e^{i{\bf q}\cdot{\bf r}'}\left\langle \boldsymbol{\xi}^{T}({\bf r},s)\boldsymbol{\xi}^{T}({\bf r}',s')\right\rangle \cdot{\bf q}\nonumber \\
= & 2\mu_{b}T\delta(s-s')(2\pi)^{d}\delta({\bf k}+{\bf q}){\bf k}\cdot{\bf I}\cdot{\bf q}\nonumber \\
= & -2\mu_{b}Tk^{2}\delta(s-s')(2\pi)^{d}\delta({\bf k}+{\bf q})
\end{align}
this demands ${\bf k}=-{\bf q}$. Similarly, we also have
\begin{equation}
\left\langle {\bf k}\cdot\tilde{\boldsymbol{\xi}}_{k}^{A}(s){\bf q}\cdot\tilde{\boldsymbol{\xi}}_{q}^{A}(s')\right\rangle =\frac{D_{b}}{\tau_{b}}e^{-|s-s'|/\tau_{b}}k^{2}(2\pi)^{d}\delta({\bf k}+{\bf q})
\end{equation}
Bringing these into Eq.\eqref{eq:corrAT}, we have Eqs.\eqref{eq:corr_eta_a}
and \eqref{eq:corr_eta_t} in the main text.
\end{document}